\documentclass[twocolumn,pre,aps,superscriptaddress]{revtex4-1}

\usepackage[pdftex]{graphicx}

\begin{document}
\title{Dynamics on networks: competition of temporal and topological correlations}

\author{Oriol Artime}
\author{Jos\'e J. Ramasco}
\author{Maxi San Miguel}
\affiliation{Instituto de F\'{\i}sica Interdisciplinar y Sistemas Complejos IFISC (CSIC-UIB), Campus Universitat de les Illes Balears, 07122 Palma de Mallorca, Spain}

\begin{abstract}
Links in many real-world networks activate and deactivate in correspondence to the sporadic interactions between the elements of the system. The activation patterns may be irregular or bursty and play an important role on the dynamics of processes taking place in the network. Information or disease spreading in networks are paradigmatic examples of this situation. Besides burstiness, several correlations may appear in the process of link activation: memory effects imply temporal correlations, but also the existence of communities in the network may mediate the activation patterns of internal an external links. Here we study the competition of topological and temporal correlations in link activation and how they affect the dynamics of systems running on the network. Interestingly, both types of correlations by separate have  opposite effects: one (topological) delays the dynamics of processes on the network, while the other (temporal) accelerates it. When they occur together, our results show that the direction and intensity of the final outcome depends on the competition in a non trivial way.
\end{abstract}


\maketitle

\section{Introduction}

Networks are used to represent or establish the interaction framework between the basic elements of complex systems. Several features of the network may play a significant role in the behavior of the system dynamics. Some examples are the network topology (e.g., heterogeneity in the connections, abundance of loops, multiplex structure) or the presence of different intensities (weights) in the links  \cite{newman2010networks, barrat2008dynamical,pastor2001epidemic,barrat2004,kivela2014multilayer,wang2016statistical,wang2015immunity,zhao2014immunization,diakonova2016irreducibility}, to name a few. Other than static properties, the temporal nature of the activation of the elements of the network itself can bring new phenomena in the system dynamics that do not have a counterpart in static or time-aggregated networks like reachability, causality or the time-ordering contact sequences \cite{cattuto2010,miritello2011,holme2012,gauvin2013,machens2013,karsai2014time, vestergaard2014}. The temporal dimension becomes specially important when dealing, for example, with human interactions since time inhomogeneities between consecutive activations are notable. Long-tailed distributions in the link activation inter-event times are a generic feature observed and modeled in human dynamics at the level of links  \cite{barabasi2005,oliveira2005human,vazquez2006,vazquez2007impact,gonccalves2008human,gama2009,iribarren2009impact,radicchi2009,meiss2009} and nodes \cite{fernandez2011update,perra2012}. Such unevenly distributed activations affect the propagation of information across the network and change the outcome of models running on the graph \cite{fernandez2011update, stark2008decelerating,onnela2007structure,karsai2011small,takaguchi2011,machens2013,
vestergaard2014,liu2014,delvenne2015diffusion}. One recent example is the shift detected in the epidemic threshold of the SIS and SIR models running in activity-driven temporal networks \cite{sun2015contrasting}.

\begin{figure*}
\includegraphics[width= 16cm]{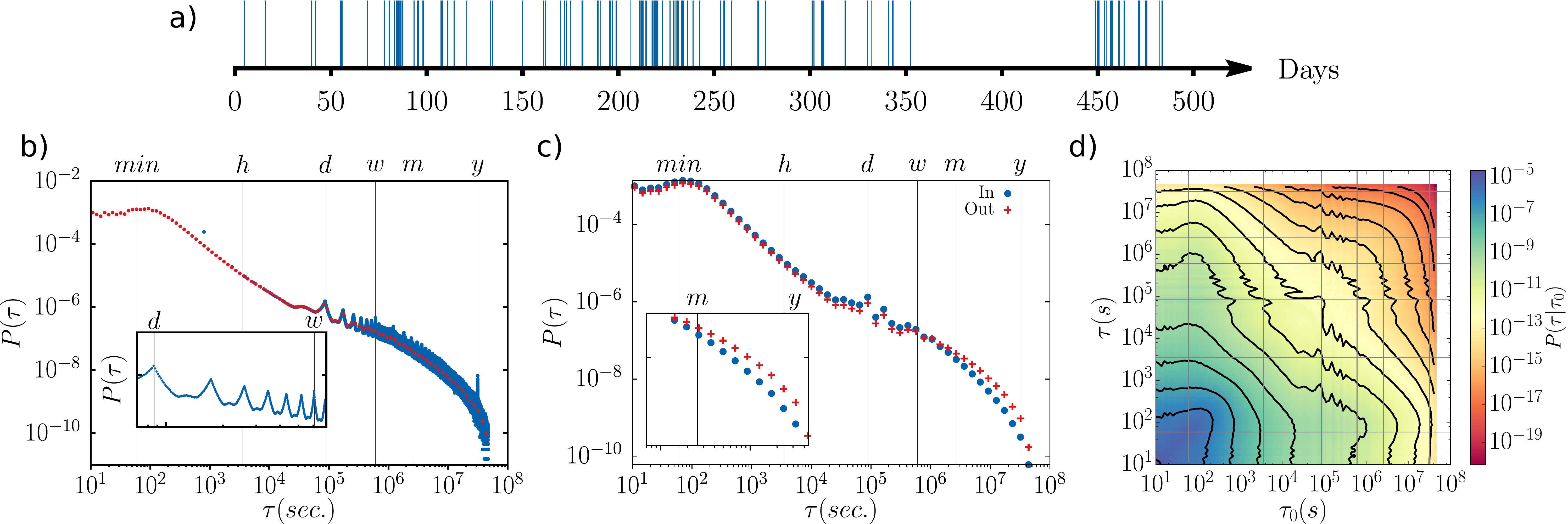}
\caption{In a), typical activity pattern for a directed link of two generic Twitter users. Each vertical tick represents a directed message. The inter-event time $ \tau $ corresponds to the separation between two consecutive ticks. In b), inter-event time distribution for all the users. Blue and red dots represent the lin- and log-binned scales in the $\tau$ axis. The localized maxima in the tail of the distribution correspond to circadian rhythms, as shown in the bottom inset. In c) the inter-event time distributions for internal and external links to the communities detected with Infomap are displayed. The inset shows the detail of the tail of the distributions. In d), colormap depicting the conditional probability of two consecutive inter-event times. Equiprobable lines are shown on top, evenly spaced in the log-scale from probabilities of $ 10^{-5} $ to $ 10^{-18} $. \label{fig:emp}}
\end{figure*}

It is extensively reported that single Poisson processes, characterized by an exponential inter-event time distribution, do not faithfully describe real interactions patterns in a variety of systems. The times between consecutive link activations are better described by heavy-tailed distributions as a reflection of the large heterogeneity at play. In first studies of this problem, the description has been restricted  to the characterization of the inter-event time distributions. This first approximation does not address the question of the correlations or independence between consecutive inter-event times. The activation inter-event times of real-world networks can show, however, a much richer behavior with different types of correlations involved \cite{goh2008burstiness,onnela2007structure,karsai2011small}. Link activations can display temporal correlations, the so-called memory effects \cite{goh2008burstiness}, in which the activations take place in avalanches keeping short inter-event times close together, followed by long periods of infrequent activity. From the point of view of the variable describing the time between consecutive activations of the same link $\tau$, this phenomenon can be understood as a positive correlation in which low values of $\tau$ are followed with high probability by low $\tau$'s and vice-versa. 
In addition, link activation can also be related to topological characteristics of the network. The simplest topologically induced inter-event time correlations appear between links connecting to the same node \cite{fernandez2011update,perra2012,liu2014}. For instance, the limited cognitive capacity of the individuals (nodes) in social networks reflects in the time and effort invested in each relation (link) impacting thus the inter-event times between consecutive contacts \cite{miritello2013}. More elaborated  correlations may rise in networks with community structure where the inter-event time activation in links internal and external to the communities can be different. In this work, we will focus on the effects of the competition between the temporal correlations and this latter topologically (community) induced correlations in link activation.

Understanding the interplay between dynamical process and temporal and structural properties of the networks has been an active area of research. In general, it has been reported that non-Poissonian inter-event time distributions, as well as temporal and topological correlations, contribute to slow down the dynamical processes on networks, especially when the dynamics on empirical networks is compared with that on randomized null models \cite{onnela2007structure,vazquez2007impact,iribarren2009impact,toivonen2009broad,karsai2011small,takaguchi2011, karsai2014time}. Some recent analytical results point to the existence of regimes where an acceleration of the dynamical process with respect to these null models might be possible \cite{rosvall2014memory, scholtes2014causality,scholtes2016higher}. It has been also pointed out the importance of different contributions to the evolution of both the network and the processes on top of it \cite{ubaldi2016burstiness}. For instance, it has been found that the absence of memory in the pathways can overestimate the time to reach equilibrium in diffusion \cite{lambiotte2015effect}. Here we approach the problem from a new perspective. In order to study how dynamical processes on the network are affected by topologically induced and temporal correlations, we add them artificially with the possibility of tuning their strength. We perform exhaustive numerical simulations to discriminate the outcomes of the models in function of the power of the correlation and to grasp the mechanism behind the alterations seen in the dynamics. As paradigmatic examples we consider two dynamical processes of diverse nature: Susceptible-Infected SI spreading and the voter model as prototypes of epidemics and opinion models. Our results show that the effect of both types of correlations alone on the dynamics are opposite (memory effects accelerate the dynamics, while correlations induced by community structure slows it down). However, when both types are combined the final dynamics crucially depends on how the combination is built reflecting the competition between correlations.

\section{Methods}

\subsection{Twitter data description}

To illustrate the different types of correlations, we use a data sampling extracted from Twitter. Our dataset contains $ 73\,405\,100 $ directed tweets (replies and mentions) obtained following the activity of $ 2\,590\,459 $ unique users. In this case, the nodes of the network are the users and a directed link is established every time that a user sends a message with a defined target (replies). In total, there are $5\,812\,089$ links among the considered users. Figure \ref{fig:emp}a shows the activity pattern of a particular link. The x-axis represents time and a blue tick is displayed every time an unidirectional message is interchanged between the same pair of users. As can be seen, the events are concentrated in avalanches with periods of high activity and short inter-event times followed by long periods of inactivity. 
In the Figure \ref{fig:emp}b, the inter-event time distribution between consecutive messages of every pair of connected users is displayed. The peaks correspond to circadian rhythms of multiples of 24 hours (see inset). 

The network can be divided in groups using a community detection algorithm. In this case, we employed Infomap \cite{rosvall2008}. With this information at hand, we compute the activation inter-event times of the links internal to the  communities, $P_{in} (\tau)$, and those connecting different communities $P_{out} (\tau)$. They look similar at first sight (see Figure \ref{fig:emp}c). Nevertheless, taking a closer look at the tails, we see that $ P_{out}(\tau) $ decays slower than $ P_{in}(\tau) $. This means that the communication of members within the same community tend to be more frequent than those between members of different communities. To check the level of these differences, we apply a Kolmogorov-Smirnov test with the null hypothesis that both histograms are coming from the same distribution. The test completely rules out this possibility with certainty within our numerical precision.

Finally, the Figure \ref{fig:emp}d depicts the single link inter-event time correlations. To estimate this, we measure the conditional probability $P(\tau | \tau' )$ of a link to present an inter-event time $\tau$ after another $\tau'$. This probability shows the presence of temporal correlations in the empirical data. The Pearson auto-correlation coefficient for empirical inter-event times of the same link is $0.115$, while reshuffling the inter-event times this correlation becomes of the order of $10^{-5}$. The temporal correlations are thus low, but significant and positive.

\subsection{Modeling link activation with temporal and topological correlations}

The empirical observations offer already some insights on which features of link activation must be included to model these networks. In particular, we want to consider interactions that are distributed according to an arbitrary function (usually long-tailed), as well as taking into account the first-order correlations of their inter-event times. In addition, we want also to include  communities in order to incorporate the different paces of activation of the internal and external links. 

The first step is to build the network. We use Lancichinetti's LFR benchmark \cite{lancichinetti2008benchmark} to generate networks with planted communities.
This algorithm produces scale-free networks with a distribution of community sizes that falls as a power-law with exponent $-1$ and it has been used as a test-bed for community detection methods. The heterogeneity in the community size is important because this is is a common feature to a number of real networks. 
The mixing parameter $\mu$ controls the fraction of internal versus external links in the communities in such a way that for $\mu = 1$ the links are all external and for $\mu = 0$  are all internal, while the community structure is lost at an intermediate value of $\mu$.

\begin{figure*}
\includegraphics[width=17cm]{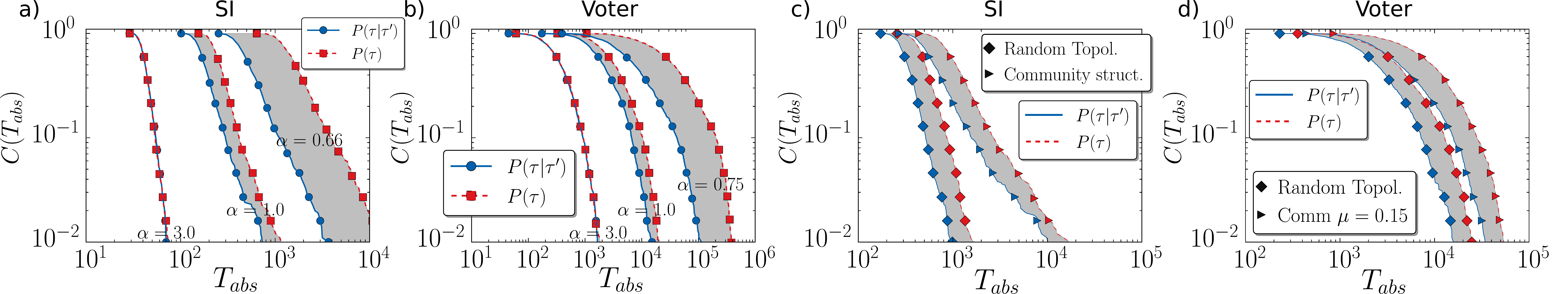}
\caption{Complementary cumulative distribution function of the times  to reach the absorbing state $C(T_{abs})$ in different topologies and with different distributions. $C(T_{abs})$ stands for the fraction of realizations still active at time $ T_{abs}$. In a), simulations performed with inter-event time distributions of equation (\ref{eq:dist}), for different values of $ \alpha $. The network is random (Erd\"os-R\'enyi) and the model is SI. The value of $\beta = 0.1$ and it is the same in all the following SI simulations. In b), the same but for the voter model. In c) (SI model) and d) (voter model) results are shown on networks with community structure ($\alpha_{in} = 1$ and $\alpha_{out} = 0.75$) and compared with those obtained on reshuffled instances of the graphs. Network sizes are $ N = 5\,000 $ for the SI simulations and $ N = 2\,000 $ for the voter simulations. In all the networks, $ \left\langle k \right\rangle = 12 $ and the results are sampled over $ 1\,000 $ realizations.  \label{fig:together2}}
\end{figure*}

We also add pure temporal correlations associated with memory in the update rule for the links. The activation rule is link-based: every link $j$ keeps a clock with its time of activation, $t_j'$, that is reset to a new time $t_j = t_j' + \tau$ after activation. When a link activates, the two nodes that it is joining interact accordingly to the dynamical model. The important question is, therefore, how $\tau$ is selected given the previous inter-event time for the same link $\tau'$. For simplicity, a Markovian stochastic extraction approach is followed so that if $\tau$ and $\tau'$ are independent  (the conditional probability is just $P(\tau|\tau') = P(\tau)$), the link activation has no memory. On the contrary, if the conditional probability $P(\tau|\tau')$ explicitly depends on both times, memory effects are present and their intensity can be tuned by changing the shape of  $P(\tau|\tau')$. To be specific, the first value of $\tau$ for every link is chosen from an independent $P(\tau)$. The ulterior update times are generated using $P(\tau|\tau')$. This is a rather general framework, the distributions can even come from empirical data as those of Figure \ref{fig:emp} (see below). Here we use the function \cite{ramasco2007transport}
\begin{equation}
\label{eq:dist}
P(\tau| \tau ' ) = \frac{P(\tau,\tau')}{P(\tau')} = \frac{(1+\alpha) \, (1+\tau')^{1+\alpha}}{(\tau+\tau')^{2+\alpha}},
\end{equation}
where the exponent $ \alpha > 0 $ is always positive. This conditional probability derives from a joint one with the simple form $P(\tau,\tau') \propto 1/(\tau+\tau')^{2+\alpha}$, which in turn leads to power-law decaying inter-event time distributions $P(\tau) \sim \tau^{-1-\alpha}$ in the large $\tau$ limit, and it induces positive correlations with the following  average inter-event time activation for a single link being $ \langle \tau \rangle = (1 + \alpha + \tau') / \alpha $ \cite{ramasco2007transport}. This is a growing function of $\tau'$ implying that large values of $\tau'$ are followed on average by large values of $\tau$ and vice-versa. Note that the strength of the temporal correlations can be controlled  through the parameter $ \alpha $: the lower $\alpha$ is, the stronger the positive correlations between $\tau$ and $\tau'$ become. As in the empirical data, if the network has community structure, instead of a single $ \alpha $ we use the same distribution of equation (\ref{eq:dist}) but with exponents  $ \alpha_{in} $ or $ \alpha_{out} $ depending on the nature of each particular link. Since generally the communication inside the same community is more frequent than across communities, we assume that $\alpha_{in} \ge \alpha_{out}$. Tuning thus $\mu$ and controlling the values of $\alpha_{in}$ and $\alpha_{out}$, one can pass from a network with strong topologically induced correlations in the links activation to a homogeneous situation with a single type of links.

When needed for comparison, we introduce null models in the network activation. These null models are build differently depending on the type of correlations. In the case of temporal correlations alone, the null model is produced by generating link activation inter-event times using $P(\tau)$ instead of $P(\tau|\tau')$. When communities are present and the inter-event time distributions can be different for internal and external links, we reshuffle the connections maintaining the node degrees and the activation statistics of the  reshuffled links to ensure that the proportion of internal and external links is kept invariant. Finally, when both type of correlations are present the inter-event times come from distributions $P(\tau|\tau')$ whose exponents depend on the nature of the link with respect to the communities. The way to generate a null model in this last case is to use the corresponding $P(\tau)$ and to reshuffle the connections preserving the link statistics. 

The models run on these networks are the Susceptible-Infected (SI) spreading \cite{daley2001epidemic} and the voter model \cite{liggett1999stochastic}. The SI is a simple model simulating the spreading of information or the early stages of an infection invading the network. The nodes can be in two states: susceptible $ S $ and infected $ I $. Infected nodes can transmit the disease to susceptible ones with probability $ \beta $ every time they interact. The final system state is always fully infected, but the time to reach it depends on $\beta$ and on the link activation dynamics. The initial condition is a single randomly chosen node infected and the rest susceptible. In the voter model, on the other hand, each node has a state, or opinion,  that can take the values $+1$ and $-1$. In every update, the nodes blindly copy the state of one of their neighbors chosen at random. The initial condition is a random configuration with equal probability for each node of being $+1$ or $-1$. The dynamics of the system depends on the effective network dimensionality \cite{suchecki2005voter}. General complex networks are high dimensional where the system remains in a long-lived dynamical state until the absorbing state (consensus) is reached due to finite size fluctuations in a time that is a function of the system size. As a first approach, we have chosen these two models because they  present very simple dynamics with a single time-scale directly related to the system size. In more involved models with further time-scales or characteristic times, the interaction between the link activation and model dynamics must be also taken into account. However, this goes beyond the scope of the present study.

\section{Results}

\subsection{Effects of isolated temporal and topological correlations}

This section reports the impact of each type of correlations alone on the model dynamics. In particular, we compare the time to arrive at the absorbing state, $T_{abs}$, when correlated link activation is considered against the corresponding uncorrelated case. For the SI model, the absorbing state corresponds to the configuration with all nodes infected, while for the voter model to the general consensus. The results of the simulations are shown in Figure \ref{fig:together2}.

We focus first on the effects of temporal correlations running the models on a random network, the so-called Erd\"os-R\'enyi graph, for several values of $\alpha$. We compare the values of $T_{abs}$ for the correlated dynamics ($P(\tau|\tau')$) with the null model ($P(\tau)$). The results are displayed in Figures \ref{fig:together2}a (SI model) and \ref{fig:together2}b (voter model). We find that for strong correlations (low $\alpha$) the outcome is an acceleration of the dynamics. In fact, the presence of temporal correlations notably reduces $T_{abs}$.  As temporal correlations become lower, the separation between the correlated and uncorrelated case is less significant. Eventually, when correlations are low there is no acceleration at all, being both cases indistinguishable. This is the case of the leftmost curves in Figures \ref{fig:together2}a and \ref{fig:together2}b, in which the correlated and uncorrelated case overlap almost perfectly. Note also that the system dynamics in the limit of large $\alpha$ tends toward the Poissonian one.

When the topologically induced correlations enter into play, the scenario becomes richer. Now, the networks to be used are generated with the LFR benchmark. The topology is characterized by a tunable community structure and, therefore, there are two types of links: internal and external, with their corresponding distributions with different exponents $\alpha_{in}$ and $\alpha_{out}$. In Figures \ref{fig:together2}c and \ref{fig:together2}d, we see that in a first analysis and regardless of the presence of temporal correlations the dynamics in networks with communities suffers a delay as already reported in \cite{onnela2007structure,karsai2011small}. In this case, the communities act as local traps for the spreading process or for the diffusion in the voter model. Moreover, on the same type of topology (either random or with communities) when comparing the dynamics with $ P(\tau|\tau') $ and $P(\tau)$ we obtain that the temporal correlated case speeds up the system with respect to the uncorrelated one. It is worthy to make a final remark in this section. Note that for the SI model (Figure \ref{fig:together2}c) the way the cumulative distributions decay is different in the random networks and in the network with communities. We obtain a much broader distribution, with a slower decay in the latter case. This reveals the fact that the SI is a model highly sensitive to the community structure. On the opposite, the cumulative curves for the voter model (Figure \ref{fig:together2}d) decay similarly, maintaining approximately the same distance and functional shape. In fact, it has been discussed in the literature that the voter dynamics is not so sensitive to community structure in undirected networks except for the limit in which the communities are sparsely connected \cite{castello2007anomalous, masuda2014voter}.

\subsection{Role of temporal backbones in the acceleration of the dynamics}

\begin{figure*}
\includegraphics[width=16cm]{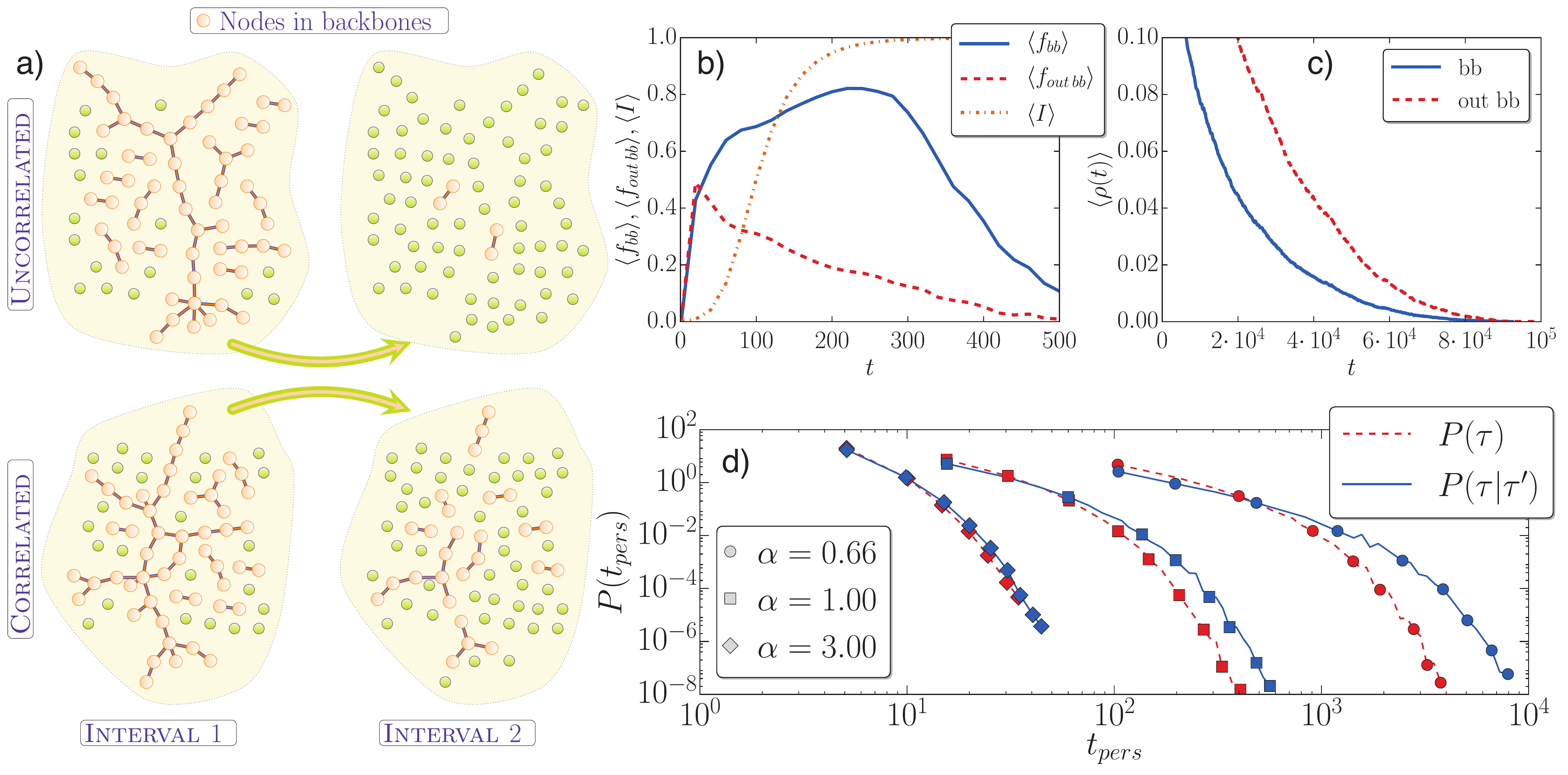}
\caption{In a), sketch of the persistence of the backbones in two consecutive intervals. The network is random and with $ 75 $ nodes, $ \langle k \rangle = 6$, and the link activation is simulated with $ \alpha = 0.2 $. In b), the average fractions of links contributing to the spreading in the backbone and outside the backbone is displayed for the SI as a function of time. In the background, the average fraction of infected nodes $\langle I \rangle$ is depicted to provide a guide on the global state of the outbreak. 
 In c), the average number of active interfaces inside and outside the backbones for the voter model. In b) and c), the averages are taken over $ 1000 $ realizations in networks of $2000$ nodes, the inter-event times are uncorrelated $P(\tau)$ with $\alpha = 0.75$.
In d), persistence probability of the consecutive time spent by a link belonging to a backbone. The time interval is different for different values of $ \alpha $ but it is the same for the correlated and uncorrelated cases. Simulations are done in Erd\"os-R\'enyi networks of $ 1000 $ nodes and $ \langle k \rangle = 12$, averaged over $ 100 $ realizations.  \label{fig:persistence}}
\end{figure*}

While the consequence of topological correlations alone is well understood \cite{castello2007anomalous,onnela2007structure,karsai2011small,karsai2014time}, the same cannot be said for the temporal correlations. Which is the mechanism that leads to the acceleration of the models' dynamics? In order to find an intuitive answer, we introduce next the concept of activation backbone. Commonly, the backbone or ``superhighway" of a weighted network refers to the set of links that act as the main skeleton through which the information travels across the system \cite{wu2006transport}. In this case, we count in a given time window how many times each link has become active. This defines a weight for every link. The backbone is calculated as the largest connected component at percolation when the links are dropped from weakest to strongest. The backbone marks the most likely pathways used in the spreading of information for any dynamical process occurring on the network. 

To compute the backbones we proceed as following. We define a time window duration, which depends on $ \alpha $ but does not depend on whether the update is correlated or not to ensure a fair comparison between both cases. The reason for this is that we want to characterize the dynamics taking place in the time scales needed to reach the absorbing state $T_{abs}$, and this is strongly dependent on $ \alpha $. Therefore, it is necessary to divide the time window in given number of equal intervals. It is important to reach a compromise between capturing the system dynamics and having good link activation statistics in every of these interval. This constraint determines the range of possible number of intervals to be used. We are using a division in $100$ intervals, although we tried with different numbers to obtain qualitatively similar results.

\begin{figure*}
\includegraphics[width=14cm]{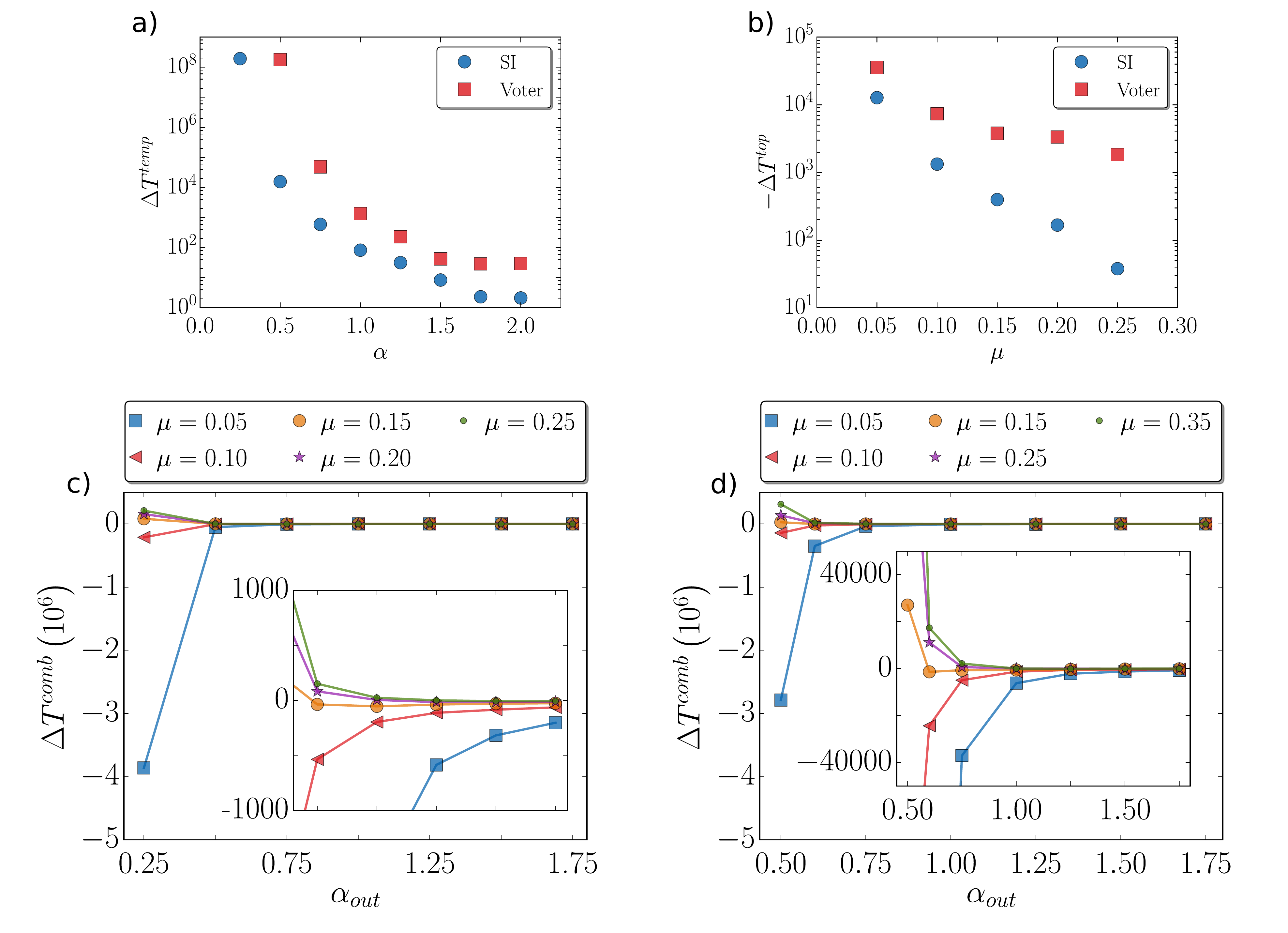}
\caption{In a), time variation $\Delta T^{temp}$ between the correlated dynamics and the uncorrelated one as a function of $\alpha$ in a random scale-free network with exponent $\gamma = -2$. In b), time variation $\Delta T^{topol}$ between a network with community structure and a null model with the links reshuffled. In the reshuffled case, the proportion of in and out links is maintained. The exponents are $\alpha_{in} = 1$ and $\alpha_{out} = 0.75$. In c) (SI) and d) (voter), for each value of $\mu$ the time variation $\Delta T^{comb}$ between the correlated dynamics and a null model without temporal correlations and with the links reshuffled but maintaining the proportion of in and out as a function of $\alpha_{out}$. $\alpha_{in} = \alpha_{out} + 0.25$. The inset is a zoom of the bifurcation area. In all cases, the network size is $N = 2\, 000$ for the SI and $1\,000$ for the voter model. Each point corresponds to an average over $ 1\,000 $ realizations.
\label{fig:mix}}
\end{figure*}

How do we relate the dynamical models with the backbones? In the case of the SI, when the disease is outside the backbone due to the infrequent link activation, the contagion process is slow and local around the neighborhood of the infected nodes. It is enough for the infection to hit a node in the backbone to spread quickly across the network. In Figure \ref{fig:persistence}b, we plot $\langle f_{bb} \rangle $ and $\langle f_{out\,bb} \rangle $, which are the average fraction of links inside and outside the backbone that participate in the spreading. At initial times when the average fraction of infected nodes $\langle I \rangle$ is very low, the spreading is equally probable to be through links of the backbone or outside the backbone. As time goes on and $\langle I \rangle$ grows, we see that the spreading contributions mainly come from those links belonging to the backbone. The case of the voter model is more convoluted since there is no limitation on the number of times a node can change its state. To shed some light on how it works in this case, we compute at the end of each time interval the fraction of interfaces (number of links joining nodes of different states) inside and outside the backbone, $\rho_{bb}$ and $\rho_{out}$ respectively (Fig. \ref{fig:persistence}c). We find that there is always more order inside the backbone than outside, so $\rho_{bb}(t) < \rho_{out}(t)$.  

The temporal behavior of the backbone depends on the presence or absence of memory effects: if $ P(\tau|\tau') $ is used in the update, the backbones are more persistent than in the case of $P(\tau)$. In Figure \ref{fig:persistence}a, a sketch of this phenomenon is displayed. The fraction of a backbone in one interval and remaining in the following one is computed. In the correlated case, a larger part of the old backbone is contained in the next one than in the uncorrelated case. It is interesting to note that the size of the backbones does not depend on the temporal correlations, since they are computed at the percolation transition and there is no topological correlations. This means that in each time interval, the backbones are formed approximately by the same number of nodes. What changes is the probability of finding the same set of links belonging to the backbone from one interval to another. Indeed, the acceleration is not produced by an increase in the size of the superhighways but by their persistence in time. We compute the time that a link spends consecutively within the backbones, $ t_{pers} $. If a link stops belonging to the backbones in a given interval, its temporal counter is reset to zero. The distributions of $t_{pers}$ are shown in Figure \ref{fig:persistence}d, where we can see that as the correlation increases, so does the persistence of a link in the backbone. 
The presence of the backbones and their temporal behavior is independent of the dynamical processes occurring on top of a network. Still, persistent backbones provide privileged pathways for the spreading of information and diffusion in the networks. For the SI, the backbones in uncorrelated networks change fast in time hindering the development of the global breakout. Similarly, for the voter model the nodes in the backbone reach consensus quicker in the correlated networks, which in turn pulls the rest of the network to the absorbing state.

\begin{figure*}
\includegraphics[width=17cm]{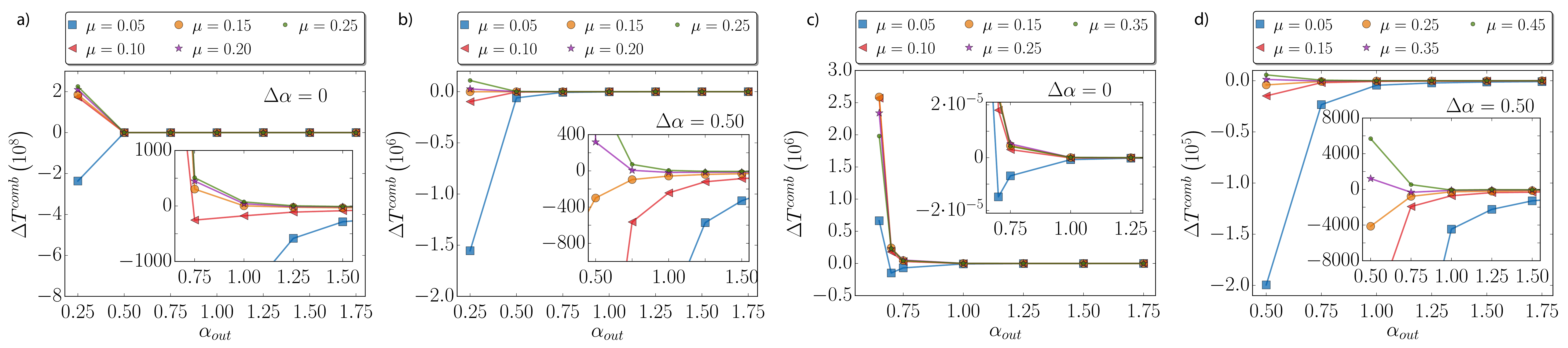}
\caption{Bifurcation diagram for several values of $ \Delta \alpha = \alpha_{out} - \alpha_{in}$. In a) and b) we show $\Delta T^{comb}$ for the SI model with the values $ \Delta \alpha = 0 $ and $ \Delta \alpha = 0.5 $, respectively. Similar plots in c) and d) for the voter model.
\label{fig:gaps}}
\end{figure*}

\subsection{Effects of mixed correlations on the dynamics}

In the discussion above some instances with mixed temporal and topologically induced correlations have been shown. In this section, the objective is to analyze systematically how the mix impacts the model dynamics. To quantify the effects of the different correlations, the variation of $T_{abs}$ with respect to the corresponding null model is depicted as a function of $\mu$ and $\alpha$  in Figure \ref{fig:mix}. The variation is measured setting a threshold in the cumulative distributions $C(T_{abs})$ (as those of Figure \ref{fig:together2}) at $0.1$ and calculating the difference between the dynamics on the null model and on the (topologically and/or temporally) correlated networks, $\Delta T = T_{abs}^{null}(0.1)-T_{abs}^{corr}(0.1)$. With this definition, $\Delta T > 0 $ implies a speeding up due to correlations, while $\Delta T < 0$ a slowing down. The first panel \ref{fig:mix}a shows how the presence of temporal correlations in the link activation alone speeds up the arrival at the absorbing state. We compare the absorbing times for the temporally correlated update $P(\tau|\tau')$ against the null model on the same random graph as a function of the parameter $\alpha$. We see that as temporal correlations get weaker, the differences in the absorbing times $\Delta T$ for both models decrease. On the contrary, the differences diverge as $\alpha \to 0$. The topologically induced correlations alone delay the dynamics as can be seen in Figure \ref{fig:mix}b for all the values of $ \alpha $. The results are obtained using fixed values of $ \alpha_{in} = 1 $ and $\alpha_{out} = 0.75$, but without temporal correlations. The presence of community structure delays the dynamics with respect to the null model and the effect is stronger when $\mu$ decreases.

The dynamics under the mix of the two types of correlations is far from trivial. Here we set a gap of $ \Delta \alpha = \alpha_{out} - \alpha_{in} = 0.25 $. As shown in Figures \ref{fig:mix}c and \ref{fig:mix}d, the system dynamics can accelerate or decelerate depending on the particular mix. When the community structure is not very defined (high $\mu$), an increase in temporal correlations with lower $\alpha$ speeds up the arrival at the absorbing state. However, for lower $\mu$ (better defined communities), the effect is the opposite and the dynamics notably slows down for small $\alpha$. There exist a bifurcation in a middle ground value of $\mu$ where the correlations exactly compensate and no effect is observed. This value is found to be $\mu \approx 0.15$ in the two studied models. 

\begin{figure*}
\includegraphics[width=15cm]{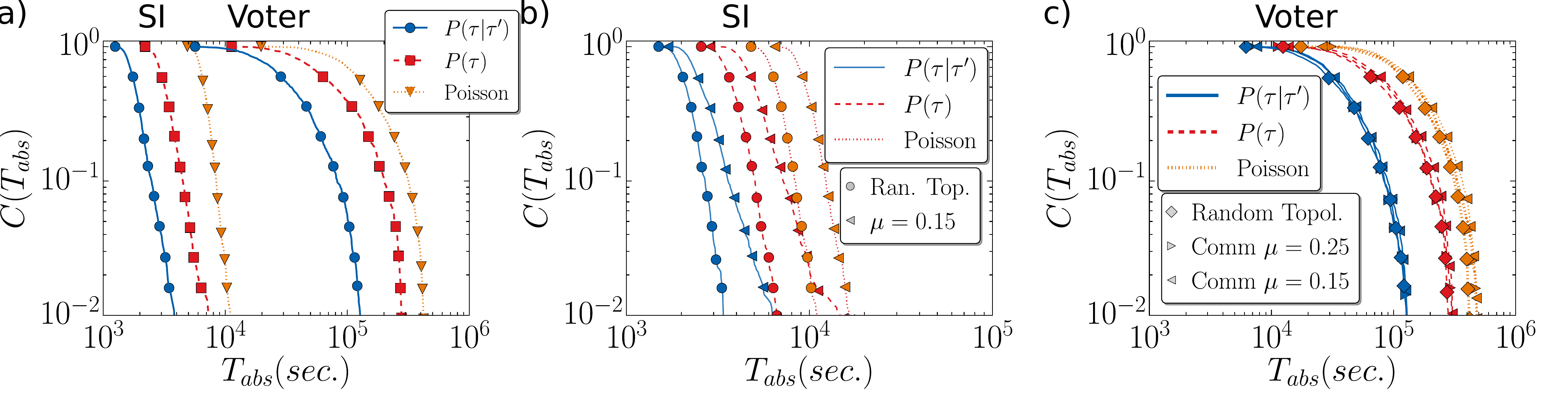}
\caption{Complementary cumulative distribution function for the times to reach the absorbing state, using empirical distributions $P_{emp}(\tau|\tau')$ in the update. The network sizes are $N = 10\, 000$ for the SI simulations and $N = 5\, 000$ for the voter model. The results were obtained running $1\, 000$ realizations. In a), effects of the temporal correlations in an Erd\"os-R\'enyi network. In b), effects of the temporal and topological correlation on the SI model dynamics. In c), the same for the voter model. In these last two, the random topology corresponds to a link reshuffling of the network with communities respecting the node degree. \label{fig:emp_simu}}
\end{figure*}

The gap value $ \Delta \alpha = 0.25 $ is arbitrary. We study next how this gap can influence the results. We first take a look at $ \Delta \alpha = 0$, i.e., $ \alpha_{out} = \alpha_{in} $. This extreme case corresponds to the case in which the nodes of the same community interact in the same way as those connecting different communities. Figures \ref{fig:gaps}a and \ref{fig:gaps}c  show the results for the SI and the voter model, respectively. In the SI model, a decrease in the value of $\mu$ where the bifurcation appears is observed. In the limit of low $ \alpha $, for $ \mu = 0.1 $ there is still an acceleration, while for $ \mu = 0.05 $ the arrival at the absorbing state is delayed. For the voter model, on the other hand, the bifurcation does not appear in the range of values of $\mu$ explored. In the extreme case of $ \mu = 0.05 $ and low $ \alpha $, it seems that the curve begins to bend down, but it finally goes to the region $\Delta T^{comb} > 0$, where the speeding up dominates.

After a zero gap, we also consider an extended difference between $ \alpha_{in} $ and  $\alpha_{out}$ with $\Delta \alpha = 0.5$. The activation of links connecting nodes in different communities is much less frequent than those in same communities in this case. In Figures \ref{fig:gaps}b and \ref{fig:gaps}d, we display the behavior of $\Delta T^{comb}$ for the SI and the voter model. The bifurcation still appears, and it occurs at higher values of $ \mu $ ($\mu \approx 0.15-0.2$ for the SI and $\mu \approx 0.3$ for the voter model). To summarize, the presence of the bifurcation occurs for different gaps between $\alpha_{in}$ and $\alpha_{out}$, although the value of $\Delta \alpha$ has an impact on the bifurcation value of $\mu$. When in- and out-links activate at same or very similar rhythm, we find that the bifurcation $\mu$ is decreased. In the voter model and for $\Delta \alpha = 0$, it decreases so much that we do not appreciate the bifurcation in the studied region of parameters. On the contrary, when links internal to the communities activate much more frequently than external links, we see that the $ \mu$ of bifurcation increases for both models.

\subsection{Empirical distributions}

In the methods section, we introduced a database of Twitter communications to illustrate the different correlations that can be present in an empirical environment. The main characteristics found in the inter-event times, which were later implemented in the model networks, were a fat-tailed distribution for $P(\tau)$ and a positive correlations between consecutive activation times as given by $ P(\tau|\tau') $ (see Figure \ref{fig:emp}). Beyond Twitter, other datasets of human recorded social interactions are reported to have similar characteristics \cite{barabasi2005,oliveira2005human,vazquez2007impact,gonccalves2008human,gama2009,radicchi2009,onnela2007structure,goh2008burstiness,karsai2011small}. 
We used these features as inspiration to construct the temporal activation patterns of our networks and we also mentioned that $P(\tau)$ and $P(\tau|\tau')$ could be generic functions including those estimated from empirical data. We show now how this is carried out with the Twitter dataset on networks with random topology or communities generated with the LFR benchmark. In this case, the inter-event times for the link activations are taken from the $P(\tau)$ and $P(\tau|\tau')$ shown in Figure \ref{fig:emp}. 

The results concerning the absorbing times $T_{abs}$ of the SI and voter model run on these networks are displayed in Figure \ref{fig:emp_simu}. The cumulative distributions $C(T_{abs})$ are plotted for uncorrelated inter-event times, for correlated ones using $P(\tau|\tau')$ and, for completeness, also for a Poissonian process with the same average inter-event time $\langle \tau \rangle$. In Figure  \ref{fig:emp_simu}, we see that the change from Poissonian to power-law distributed inter-event times accelerates the arrival at the absorbing state regardless of the presence or absence of other correlations. Furthermore, in agreement with the previous results, temporal correlations alone speed up the dynamics in random networks (see Figure \ref{fig:emp_simu}a). Topologically induced correlations are added in Figures \ref{fig:emp_simu}b and \ref{fig:emp_simu}c. The absorbing times for the SI model in Figure \ref{fig:emp_simu}b follow the patterns explained before: the community structure delay the dynamics (see the curves for $(P(\tau)$ or Poissonian inter-event times on a random networks and for $\mu = 0.15$), while the temporal correlations tend to accelerate it (blue solid curves). The voter model is not so sensitive to the community structure and, therefore, the curves obtained on random networks and for $\mu = 0.15$ and  $\mu = 0.25$ almost overlap in Figure \ref{fig:emp_simu}c. The main observable effect in this latter case is the acceleration due to temporal correlations. Although the empirical distributions $P_{in}(\tau)$ and $P_{out}(\tau)$ are different, they are closer than the theoretical curves with a gap $\Delta \alpha = 0.25$. This may explain the larger differences in $T_{abs}$ seen between the curves for correlated and uncorrelated networks in Figure \ref{fig:together2}d with respect to those in \ref{fig:emp_simu}c.

\section{Conclusions}

In this work, we have addressed the question of how temporal correlations in the link activation and their interplay with the topology  affect the evolution of models on networks. We introduce a model where the link activation correlations can be added manually and their strength tuned. Our results have been illustrated using two paradigmatic dynamical models: the SI and the voter models. By introducing tunable temporal and topologically induced correlations in the link activation, we have shown that while topologically induced correlations due to the presence of communities alone tend to slow down the dynamics on the network, temporal correlations speed it up. The origin of this acceleration is in the formation of stable backbones that allow for a faster communication of the elements of the system. In case of the SI, this brings a faster spreading in the network while in the voter model the persistence of the backbone induces the formation of ordered structures that pull the system towards consensus. With respect to topologically induced correlations and motivated by data analysis, we assume that there are communities in the networks and that the activation rate of the links inside and outside communities is not equal. When acting alone, the activation patterns induced by the communities have a delaying effect on the dynamics.

The procedure proposed to generate different activation patterns in the links is highly flexible. We have performed the analysis with a theoretical expression for the inter-event time distribution $P(\tau)$ and its conditional form $P(\tau|\tau')$. The framework is, nevertheless, general and admits any other functional forms for these distributions including those coming from empirical data. This has been illustrated with an example using Twitter interaction data. 

Finally, we have focused on how the presence of mixed correlations impacts model dynamics. We have explored different ranges of strength of temporal and topologically induced correlations. Interestingly, when both types of correlations are present, as occur in most empirical networks, the final dynamics crucially depends on the mix. Temporal correlations can accelerate or delay the dynamics depending on how defined the community structure is. We observe a community mixing parameter $\mu$ for which a bifurcation on the effect correlations on the model dynamics takes place. This point changes with the difference between the activation patterns of the links internal and external to the communities but its presence seems to be quite generic. The more similar the activation patterns of external and internal links are, the strongest has to be the community structure to find the bifurcation.    

Partial financial support has been received from the Spanish Ministry of Economy (MINECO) and FEDER (EU) under the project ESOTECOS (FIS2015-63628-C2-2-R), and from the EU Commission through the project INSIGHT.

\end{document}